\documentclass[12pt,urlcolor=black, linkcolor=black]{article}

\usepackage{amsmath, amsthm, amssymb, amsfonts, amscd, latexsym, amsbsy, bm, slashed}
\usepackage{mathtools}
\usepackage[sort, numbers]{natbib}
\usepackage{graphicx}
\usepackage[letterpaper]{geometry}
\usepackage[usenames, dvipsnames]{xcolor}
\allowdisplaybreaks[1]
\usepackage[colorlinks,citecolor=blue]{hyperref}
\usepackage[bbgreekl]{mathbbol}
\usepackage{enumitem,cleveref}
\usepackage{tikz-cd}

\begin{document}
\begin{titlepage}
\vskip 12.mm

\begin{center}
{\bf\Large{
Comments on flavor symmetry breaking and three-dimensional superconformal index
\\[6.8mm]
}}

\vskip.5cm
\large{\textbf{Ilmar Gahramanov}$^{1,2}$}

\vskip.5cm
{\small{\textit{$^1$Department of Physics, Bogazici University, 34342 Bebek, Istanbul, Türkiye}\\}}
\vskip.2cm
{\small{\textit{$^2$Center for Mathematics and its Applications, Khazar University, Baku, Azerbaijan}\\}}
\vskip.3cm
{\small{\texttt{ilmar.gahramanov@bogazici.edu.tr}}}
\end{center}
\vskip.35cm
\baselineskip 12pt
\begin{abstract}
We investigate the superconformal index of a three-dimensional $\mathcal N=2$ supersymmetric $SU(2)$ gauge theory with four chiral doublets and a monopole superpotential. This theory confines with the breaking of the global $SU(4)$ symmetry to $Sp(4)$. We obtain the index of this theory by integrating out a flavor from the $s$-confining theory with six chiral doublets. The superconformal index vanishes for generic values of the flavor fugacities. However, for a specific choice of fugacities it becomes a distribution supported on the maximal torus of the unbroken group, which allows us to describe the flavor symmetry-breaking phenomenon. The resulting identity has an interpretation as the inversion relation for an integrable lattice spin model.
\end{abstract}
\end{titlepage}

\pagenumbering{arabic}
\setcounter{page}{2}

\section{Introduction}

Supersymmetric duality relates the infrared limits of different supersymmetric gauge theories. The basic example of such duality \cite{Seiberg:1994pq} is a four-dimensional $SU(N_c)$ ``electric'' gauge theory with $N_f$ flavors of quarks in the fundamental and antifundamental representations, which possesses a dual description in terms of $N_f$ ``magnetic'' flavors of quarks charged under the $SU(N_f-N_c)$ gauge group. For $N_f = N_c$, the infrared theory is confined, and the classical constraint on the meson and baryon fields is modified by a one-instanton effect, leading to chiral symmetry breaking. In the case $N_c=2$ the global symmetry $SU(4)$ is broken\footnote{The same happens for the two-color non-supersymmetric QCD with two flavors, which enjoys the extended chiral symmetry $SU(4)$ instead of $SU(2)\times SU(2)$, see e.g. \cite{Smilga:1994tb,Kogut:1999iv}, and \cite{Magnea:1999iv,Dunne:2002vb} for the three-dimensional case.} to its maximal vector-like subgroup $Sp(4)$ by the chiral condensate \cite{Seiberg:1994bz}. The chiral symmetry breaking for four-dimensional ${\mathcal N}=1$ supersymmetric gauge theories was considered on the level of superconformal indices in \cite{Spiridonov:2014cxa} by Spiridonov and Vartanov.

Three-dimensional $\mathcal N=2$ supersymmetric gauge theories have rich structures \cite{Intriligator:1996ex,deBoer:1996mp,Aharony:1997bx,deBoer:1997kr,Amariti:2016kat}, such as duality, confinement, and chiral symmetry breaking, similar to four-dimensional $\mathcal N=1$ supersymmetric gauge theories\footnote{The three-dimensional $\mathcal N=2$ theory has the same number of supercharges as a four-dimensional $\mathcal N=1$ theory. In fact, the three-dimensional $\mathcal N=2$ theories discussed in this work are the dimensional reduction of the corresponding four-dimensional $\mathcal N=1$ theories \cite{Dolan:2011rp,Spiridonov:2014cxa, Aharony:2013dha, Amariti:2015xna}.}. In this work, we consider the flavor symmetry-breaking phenomenon for the three-dimensional $\mathcal N=2$ $SU(2)$ theory with $N_f=4$ chiral doublets\footnote{Throughout the paper $N_f$ denotes the number of chiral multiplets in the fundamental (doublet) representation of $SU(2)$. In this counting, the four-dimensional case $N_f=N_c=2$ discussed above corresponds to $N_f=4$, and the $s$-confining theory with three flavors corresponds to $N_f=6$.} by using the superconformal index technique. Similar to the case of the three-dimensional $S_b^3$ partition function \cite{Spiridonov:2014cxa}, the superconformal index of such a theory involves Dirac delta functions which reflect the presence of symmetry breaking. As the main result of the paper one may regard the following identity
\begin{align} \nonumber
		\sum_{m \in {\mathbb Z}} \oint \frac{(q^{1+\frac{n_\alpha\pm n_w\pm m}{2}} \alpha^{-1} w^{\mp} z^{\mp};q)_\infty}{(q^{\frac{n_\alpha \pm n_w \pm m}{2}} \alpha w^\pm z^\pm;q)_\infty}  \frac{(q^{1+\frac{-n_\alpha\pm n_y\pm m}{2}} \alpha y^{\mp} z^{\mp};q)_\infty}{(q^{\frac{-n_\alpha \pm n_y \pm m}{2}} \alpha^{-1} y^\pm z^\pm;q)_\infty} \frac{(1-q^m z^{\pm 2})}{q^m z^{4m}} \frac{dz}{4\pi i z} \\ \label{invrel}
		= \frac{(q^{1\pm n_\alpha} \alpha^{\mp 2};q)_\infty}{(q^{\pm n_\alpha} \alpha^{\pm 2};q)_\infty} \frac{(q^{1 \pm n_w} w^{ \mp 2};q)_\infty}{(q^{\pm n_w} w^{\pm 2};q)_\infty} \left(\delta(\theta-\chi) \delta_{n_y,n_w}+\delta(\theta+\chi)\delta_{n_y,-n_w}\right) \;,
\end{align}
where $y=e^{2\pi\textup{i} \theta}$, $w=e^{2\pi\textup{i} \chi}$, $\delta(\theta)$ is the periodic Dirac delta function, and $\delta_{i,j}$ is the Kronecker delta. We use the usual $q$-Pochhammer symbol $(z;q)_\infty=\prod_{i=0}^{\infty} (1-zq^i)$ and the shorthand notation\footnote{ More generally, a factor containing $\pm$ signs stands for the product over all choices of signs, where the first $\pm$ is correlated with the first $\mp$, the second $\pm$ with the second $\mp$, etc.} $f(z^\pm)=f(z) f(z^{-1})$.

On the left-hand side of equality (\ref{invrel}) we have the three-dimensional $\mathcal N=2$ superconformal index of the theory with $SU(2)$ gauge group and $N_f=4$ chiral doublets. The flavor fugacities are parametrized in such a way that only the subgroup $SU(2)_w\times SU(2)_y\times U(1)_\alpha$ of the flavor group $SU(4)$ is manifest. The index has non-zero support only on the subset $y=w^{\pm1}$, $n_y=\pm n_w$ of flavor fugacities and fluxes. On this subset the four fugacities of the doublets become $\{\alpha w, \alpha w^{-1}, \alpha^{-1}w, \alpha^{-1}w^{-1}\}$, i.e. they come in inverse pairs and parametrize the maximal torus of $Sp(4)$. This is precisely the manifestation of the chiral symmetry breaking $SU(4)\to Sp(4)$ in the confining theory, similar to the four-dimensional case and to the three-dimensional squashed-sphere partition functions \cite{Spiridonov:2014cxa}. Note that the expression (\ref{invrel}) plays the role of the local inversion relation in integrable lattice models \cite{Gahramanov:2015cva,Gahramanov:2016ilb,Kels:2015bda}. One can consider the derivation of (\ref{invrel}) as another algebraic approach to the inversion relation.

The rest of this paper is organized as follows. In the remaining part of this section, we briefly recall the superconformal index and the duality for the $s$-confining $SU(2)$ theory with $N_f=6$ chiral doublets. In section \ref{sec:breaking}, we reduce the number of flavors and derive the identity (\ref{invrel}). We conclude with a few remarks and open problems in the last section.

\subsection{Superconformal index}

We consider the supersymmetric index of a three-dimensional ${\mathcal N}=2$ superconformal field theory\footnote{For a more detailed introduction, see e.g. \cite{Krattenthaler:2011da,Gahramanov:2015cva,Gahramanov:2016wxi}.} which can be defined as a partition function on $S^2 \times S^1$ with periodic boundary conditions of the fields along $S^1$ \cite{Bhattacharya:2008zy,Bhattacharya:2008bja,Kim:2009wb,Imamura:2011su}
	\begin{equation}
		{I}(q,\{ a_i \})=\text{Tr} \left[ (-1)^\text{F} e^{-\beta\{Q, Q^\dagger \}} q^{\frac12 (\Delta+j_3)}\prod_i
		a_i^{F_i} \right] \;,
	\end{equation}
where the trace is taken over the Hilbert space of the theory on $S^2$. Here $Q$ is a certain supercharge of the three-dimensional ${\mathcal N}=2$ superconformal algebra which satisfies the anti-commutation relation $\{Q, Q^{\dagger}\}=\Delta-R-j_3$, where $\Delta$ is the energy, $j_3$ is the third component of the angular momentum on $S^2$, and $R$ is the R-charge. The operator $\text{F}$ plays the role of the fermion number, $F_i$ are the charges of the global symmetries commuting with the supercharge $Q$, and $a_i$ are the corresponding fugacities.

The superconformal index can be computed exactly by the supersymmetric localization technique \cite{Willett:2016adv}, and it takes the form of the following matrix integral\footnote{We skip the contribution of the Chern--Simons term, since in this work we consider theories without this term.}
	\begin{equation}
		I(q, \{a_j\}, \{n_j\}) = \sum_{m_i}\frac{1}{|W_m|} \oint \prod_{i=1}^{\text{rank}G} \frac{dz_i}{2 \pi i z_i} Z_{\text{gauge}} (z_i,m_i; q) Z_{\text{chiral}}(z_i,m_i; a_j, n_j; q) \;.
	\end{equation}
Here $Z_{\text{gauge}}$ and $Z_{\text{chiral}}$ stand for the contributions from the vector and chiral multiplets, respectively, $|W_m|$ is the order of the Weyl group of the subgroup of $G$ preserved by the monopole flux $m$, the sum is over monopole configurations, and the integral is over the gauge fugacities. The form of the index is completely determined if one knows the group-theoretical data, i.e. the symmetries and the corresponding representations of a theory. The reader interested in the precise formulation of the three-dimensional superconformal index can refer to the papers \cite{Imamura:2011su,Gahramanov:2016wxi}.

\subsection{\texorpdfstring{Duality for 3d $\mathcal N = 2$ $SU(2)$ SQCD with $N_f = 6$ chiral doublets}{Duality for 3d N=2 SU(2) SQCD with Nf=6 chiral doublets}}

As an example, we will study the global symmetry breaking in the three-dimensional $\mathcal{N}=2$ supersymmetric QCD with $SU(2)$ gauge group and $N_f=4$ chiral doublets. To obtain such a theory, one can first take the $s$-confining theory\footnote{Three-dimensional $s$-confining theories were discussed in several works, e.g. \cite{Csaki:2014cwa,Nii:2018erm,Nii:2019ebv,Amariti:2022wae}, and the superconformal indices for such theories were presented in \cite{Gahramanov:2013xsa,Gahramanov:2016wxi}.} with the gauge group $SU(2)$ and $N_f=6$ chiral doublets. In the case of the $SU(2)$ gauge group, quarks and antiquarks transform in the same way under the action of the gauge group, i.e. the fundamental and antifundamental representations are equivalent, and therefore the flavor group is enhanced to $SU(6)$. The superconformal index of the theory can be expressed in the following form
	\begin{equation} \label{electric}
		I_{E} \ = \ \sum_{m=-\infty}^{\infty}  \oint \prod_{i=1}^6 \frac{(q^{1+(m+n_i)/2}/a_iz,q^{1+(n_i-m)/2} z/a_i ;q)_\infty}{(q^{(m+n_i)/2}a_iz,q^{(n_i-m)/2} a_i/z ;q)_\infty} \frac{(1-q^m z^2)(1-q^m z^{-2})}{q^m z^{6m}} \frac{dz}{4\pi i z}
	\end{equation}
	with the balancing conditions
	\begin{equation} \label{balcon}
		\prod_{i=1}^6 a_i = q \;\; \text{and} \;\; \sum_{i=1}^6 n_i =  0 \;.
	\end{equation}
The variables $a_i$ and $z$ are the fugacities associated with the $SU(6)$ flavor group and the $SU(2)$ gauge group, respectively. The balancing condition reflects the presence of the monopole superpotential $W=\eta Y$, which is generated when the four-dimensional theory is compactified on a circle \cite{Aharony:2013dha}. This superpotential breaks the axial $U(1)$ symmetry, and it is responsible for the symmetry breaking discussed below. In \cite{Dimofte:2012pd}, it was shown that the superconformal index of this theory has an extended $SO(12)$ flavor symmetry when coupled to four-dimensional multiplets with specific boundary conditions.

It is known that there is a magnetic dual theory without gauge degrees of freedom and with fifteen chiral multiplets in the totally antisymmetric tensor representation of the same flavor group. The superconformal index of the magnetic theory is
	\begin{equation}
		I_M  = \frac {1}{ \prod_{i=1}^6 a_i^{n_i}}  \prod_{1 \leq i<j  \leq 6} \frac {(q^{1+(n_i+n_j)/2}/a_i a_j;q)_\infty}{(q^{(n_i+n_j)/2}a_i a_j;q)_\infty} \;.
	\end{equation}

Since these theories are IR dual, the superconformal indices need to be in agreement \cite{Spiridonov:2008zr, Gahramanov:2016wxi}. The result is given by the following $q$-hypergeometric identity
	\begin{align} \nonumber
		&  \sum_{m=-\infty}^{\infty}  \oint \prod_{i=1}^6 \frac{(q^{1+(m+n_i)/2}/a_iz,q^{1+(n_i-m)/2} z/a_i ;q)_\infty}{(q^{(m+n_i)/2}a_iz,q^{(n_i-m)/2} a_i/z ;q)_\infty} \frac{(1-q^m z^{\pm 2})}{q^m z^{6m}} \frac{dz}{4\pi i z} \\ \label{mainint}
		& \qquad \qquad \qquad \qquad =  \frac {1}{ \prod_{i=1}^6 a_i^{n_i}}  \prod_{1 \leq i<j  \leq 6} \frac {(q^{1+(n_i+n_j)/2}/a_i a_j;q)_\infty}{(q^{(n_i+n_j)/2}a_i a_j;q)_\infty}
	\end{align}
with the balancing conditions (\ref{balcon}). This integral identity was proven in \cite{Gahramanov:2016wxi} and studied in \cite{Gahramanov:2016wxi,Gahramanov:2015cva,Rosengren:2016mnw}.

The identity (\ref{mainint}) can be written as a star-triangle relation (the non-operator form of the Yang-Baxter equation) \cite{Gahramanov:2015cva,Kels:2015bda}, which is a sufficient condition for integrability of Ising-type lattice spin models. The Bailey pairs and the operator form of the Yang-Baxter equation related to this integral identity were also constructed in \cite{Gahramanov:2015cva} (see also \cite{Catak:2022glx}).

Note that the basic hypergeometric integral identity (\ref{mainint}) belongs to a relatively new class of identities which needs detailed study and classification \cite{Schlosser:2017ixz}. Such identities (containing summation and integration) have recently appeared in supersymmetric gauge theory computations \cite{Krattenthaler:2011da,Kapustin:2011jm,Aharony:2013dha,Gahramanov:2013rda,Gahramanov:2014ona, Gahramanov:2015tta,Gahramanov:2016wxi,Rosengren:2016mnw,Gahramanov:2021pgu,Bozkurt:2018xno} and in integrable models of statistical mechanics \cite{Gahramanov:2015cva,Gahramanov:2017ysd,Gahramanov:2017idz,Eren:2019ibl, Catak:2021coz, Catak:2022glx}.

\section{Breaking of flavor symmetry} \label{sec:breaking}

Our goal is to reduce the number of chiral doublets from $N_f=6$ to $N_f=4$ in the integral identity (\ref{mainint}). One can perform the reduction by taking the following limit\footnote{Note that in the work \cite{Gahramanov:2013xsa}, the authors take another limit $a_5 a_6=q^{1/2}$ and obtain the $s$-confining theory without breaking the $SU(4)$ flavor symmetry. In \cite{Gahramanov:2015cva}, the authors considered superconformal indices of dual theories with $SU(4)$ flavor symmetry without a superpotential term (without the balancing condition).}
	\begin{equation} \label{limit}
		a_5 a_6  =  q, \;\; n_5+n_6=0 \;.
	\end{equation}
In this limit the meson $Q_5Q_6$ has R-charge two and is neutral under the flavor symmetries, so that the mass term for the fifth and sixth doublets is allowed. Generically this limit gives zero on both sides, however for special choices of the fugacities $a_i$ one can deal with such integrals. Let us denote
	\begin{equation}\label{pars1}
		a_1=\alpha w, \quad a_2=\alpha w^{-1},\quad
		a_3=\beta y,\quad  a_4=\beta y^{-1},
	\end{equation}
	and the corresponding discrete parameters
	\begin{equation}\label{pars2}
		n_1=n_\alpha+n_w, \quad n_2=n_\alpha- n_{w},\quad
		n_3=n_\beta+n_y,\quad  n_4=n_\beta-n_{y}.
	\end{equation}
In terms of the new variables, the balancing conditions read as
	\begin{equation}
		a_5 a_6 \alpha^2 \beta^2 = q \;, \;\;\; 2n_\alpha+ 2 n_\beta+n_5+n_6=0 \;.
	\end{equation}
Therefore the limit (\ref{limit}) corresponds to $\alpha\beta \to 1$, $n_\alpha+n_\beta\to 0$. Before taking this limit, we keep $\alpha\beta$ generic and $a_5a_6=q/(\alpha\beta)^2$.

To proceed further we consider the following integral: multiply the integral (\ref{electric}) by a function $f(y)$ holomorphic near the unit circle and integrate over the variable $y$. Then using the integral identity (\ref{mainint}) we obtain the following expression\footnote{Here we assume that $\alpha^2 \neq 1$.}
	\begin{align} \nonumber
		\int_{\mathbb{T}} &
		f(y)  I_E \frac{dy}{2 \pi \textup{i} y}
		 = \alpha^{-2n_\alpha} w^{-2n_w} \beta^{-2n_\beta}  a_5^{-n_5} a_6^{-n_6} \frac{(q^{1+n_\alpha} \alpha^{-2};q)_\infty}{(q^{n_\alpha} \alpha^2;q)_\infty} \frac{(q^{1+n_\beta} \beta^{-2};q)_\infty}{(q^{n_\beta} \beta^2;q)_\infty} \\ \nonumber
		& \quad \times  \frac{(q^{1+\frac{n_5+n_6}{2}} (a_5 a_6)^{-1};q)_\infty}{(q^{\frac{n_5+n_6}{2}} a_5 a_6;q)_\infty}
		\prod_{j=5}^{6} \frac{(q^{1+\frac{n_\alpha \pm n_w+n_j}{2}} (\alpha w^{\pm 1} a_j)^{-1};q)_\infty}{(q^{\frac{n_\alpha \pm n_w+n_j}{2}} \alpha w^{\pm 1} a_j;q)_\infty} \\
		& \quad \times
		\int_{\mathbb{T}} \frac{dy}{2 \pi \textup{i} y} f(y) \; y^{-2n_y} \; \frac{(q^{1+\frac{n_\alpha+n_\beta \pm n_w \pm n_y}{2}} (\alpha \beta w^{\pm 1} y^{\pm 1})^{-1};q)_\infty}{(q^{\frac{n_\alpha+ n_\beta \pm n_w \pm n_y}{2}} \alpha \beta w^{\pm 1} y^{\pm 1};q)_\infty}   \prod_{j=5}^{6} \frac{(q^{1+\frac{n_\beta \pm n_y+n_j}{2}} (\beta y^{\pm 1}  a_j)^{-1};q)_\infty}{(q^{\frac{n_\beta \pm n_y+n_j}{2}} \beta y^{\pm 1} a_j;q)_\infty} \;. \label{fyint}
	\end{align}
The first factor in the integrand of (\ref{fyint}) has the following sequences of poles
	\begin{align}
		y_{in} & = \alpha \beta
		w^{\pm1} q^{\frac{n_\alpha+n_\beta \pm n_w-n_y}{2}} q^{k}, \\
		y_{out} & = \frac{1}{\alpha \beta} w^{\pm1} q^{\frac{-n_\alpha-n_\beta \pm n_w-n_y}{2}} q^{-k},
	\end{align}
where $k \in \mathbb{Z}_{\geq 0}$, and the \textit{in}-poles converge to zero $y=0$ while the \textit{out}-poles go to infinity\footnote{Strictly speaking, when the exponent of $q$ in front of $q^k$ is negative, the first few poles are cancelled by zeros of the numerator. Using the identity (\ref{identity}) below, one can see that the actual poles are located at $\alpha\beta w^{\pm1}q^{|n_\alpha+n_\beta\pm n_w-n_y|/2+k}$ and $(\alpha\beta)^{-1}w^{\pm1}q^{-|n_\alpha+n_\beta\mp n_w+n_y|/2-k}$. Accordingly, the residues (\ref{resw}) and (\ref{reswinv}) vanish automatically when the corresponding exponent is negative, due to the factors $(q^{1+n_\alpha+n_\beta\pm n_w-n_y};q)_\infty$.}. In order to calculate the integral (\ref{fyint}) we deform the contour $\mathbb T$ into a contour $\mathcal C$ which separates the first in-poles ($k=0$) from the rest of the in-poles, such that there are no singularities lying on $\mathcal C$. Then the residue calculation leads to the result
\begin{align} \nonumber
		\int_{\mathbb{T}}
		& f(y)I_E \frac{dy}{2 \pi \textup{i} y}
		=\alpha^{-2n_\alpha} w^{-2n_w} \beta^{-2n_\beta}  a_5^{-n_5} a_6^{-n_6} \frac{(q^{1+n_\alpha} \alpha^{-2};q)_\infty}{(q^{n_\alpha} \alpha^2;q)_\infty} \frac{(q^{1+n_\beta} \beta^{-2};q)_\infty}{(q^{n_\beta} \beta^2;q)_\infty} \\ \label{residues}
		& \times \frac{(q^{1+\frac{n_5+n_6}{2}} (a_5 a_6)^{-1};q)_\infty}{(q^{\frac{n_5+n_6}{2}} a_5 a_6;q)_\infty} \prod_{j=5}^{6} \frac{(q^{1+\frac{n_\alpha \pm n_w+n_j}{2}} (\alpha w^{\pm 1} a_j)^{-1};q)_\infty}{(q^{\frac{n_\alpha \pm n_w+n_j}{2}} \alpha w^{\pm 1} a_j;q)_\infty}  \left( J(w) + J(w^{-1}) +J(\mathcal C) \right),
\end{align}
where
\begin{align} \nonumber
 J(w) & =    f\left(\alpha\beta w q^{\frac{n_\alpha+n_\beta+n_w-n_y}{2}}\right) \; (\alpha\beta w q^{\frac{n_\alpha+n_\beta+n_w-n_y}{2}})^{-2n_y} \;
	 \frac{(q^{1+n_y}  (\alpha \beta w)^{-2};q)_\infty}
		{(q^{n_\alpha+ n_\beta + n_w } (\alpha \beta w)^2;q)_\infty}\\ \nonumber
  	& \quad \times	\frac{(q^{1+n_\alpha+n_\beta+n_w-n_y};q)_\infty}
		{(q;q)_\infty}
		\frac{(q^{1-n_w+n_y}  (\alpha \beta)^{-2};q)_\infty}
		{(q^{n_\alpha+ n_\beta } (\alpha \beta)^2;q)_\infty}
		\frac{(q^{1+n_\alpha+n_\beta-n_y}  w^2;q)_\infty}
		{(q^{- n_w}  w^{-2};q)_\infty} \\ \label{resw}
	& \quad \times	\prod_{i=5}^6
		\frac{(q^{1+\frac{-n_\alpha-n_w+2n_y+n_i}{2}}  (\alpha \beta^2 w a_i)^{-1};q)_\infty}
		{(q^{\frac{2n_\beta+ n_\alpha+n_w+n_i}{2} } \alpha \beta^2 w a_i ;q)_\infty}
		\frac{(q^{1+\frac{2n_\beta+n_\alpha+n_w+n_i-2n_y}{2} }  \alpha w a_i^{-1};q)_\infty}
		{(q^{\frac{- n_\alpha-n_w+n_i}{2}}   \alpha^{-1} w^{-1} a_i;q)_\infty} \;,
\end{align}
\begin{align} \nonumber
 J(w^{-1}) & =  f\left(\alpha\beta w^{-1} q^{\frac{n_\alpha+n_\beta-n_w-n_y}{2}}\right) (\alpha\beta w^{-1} q^{\frac{n_\alpha+n_\beta-n_w-n_y}{2}})^{-2n_y}  \frac{(q^{1+n_y}  (\alpha \beta w^{-1})^{-2};q)_\infty}
		{(q^{n_\alpha+ n_\beta - n_w} (\alpha \beta w^{-1})^2;q)_\infty} \\ \nonumber
		& \quad \times
		\frac{(q^{1+n_\alpha+n_\beta-n_w-n_y};q)_\infty}
		{(q;q)_\infty}
		\frac{(q^{1+n_w+n_y}  (\alpha \beta)^{-2};q)_\infty} 		{(q^{n_\alpha+ n_\beta} (\alpha \beta)^2;q)_\infty}
		\frac{(q^{1+n_\alpha+n_\beta-n_y} w^{-2};q)_\infty}
		{(q^{n_w} w^2;q)_\infty}  \\ \label{reswinv}
		& \quad \times 		\prod_{i=5}^6
		\frac{(q^{1+\frac{-n_\alpha+n_w+2n_y+n_i}{2}}  (\alpha \beta^2 w^{-1} a_i)^{-1};q)_\infty}
		{(q^{\frac{2n_\beta+ n_\alpha-n_w+n_i}{2} } \alpha \beta^2 w^{-1} a_i ;q)_\infty}
		\frac{(q^{1+\frac{2n_\beta+n_\alpha-n_w+n_i-2n_y}{2} }  \alpha w^{-1} a_i^{-1};q)_\infty}
		{(q^{\frac{-n_\alpha+n_w+n_i}{2}}   \alpha^{-1} w a_i;q)_\infty} \;,
  \end{align}
\begin{align}
 J(\mathcal C) & =  \int_{\mathcal C}  \frac{dy}{2 \pi \textup{i} y} f(y) y^{-2n_y} \frac{(q^{1+\frac{n_\alpha+n_\beta \pm n_w \pm n_y}{2}} (\alpha \beta w^{\pm 1} y^{\pm 1})^{-1};q)_\infty}{(q^{\frac{n_\alpha+ n_\beta \pm n_w \pm n_y}{2}} \alpha \beta w^{\pm 1} y^{\pm 1};q)_\infty}
		\prod_{i=5}^6 \frac{(q^{1+\frac{n_\beta \pm n_y+n_i}{2}} (\beta y^{\pm 1} a_i)^{-1};q)_\infty}{(q^{\frac{n_\beta \pm n_y+n_i}{2}} \beta y^{\pm 1} a_i;q)_\infty} \;. \label{intC}
\end{align}

Now let us take the limit $\alpha\beta\to 1$ and $n_\alpha+n_\beta \to 0$. In this limit the prefactor in (\ref{residues}) vanishes, since $(q^{1+\frac{n_5+n_6}{2}} (a_5 a_6)^{-1};q)_\infty \to (1;q)_\infty=0$. The integral $J(\mathcal C)$ stays finite, since the contour $\mathcal C$ is kept away from the poles, and therefore its contribution disappears. On the other hand, the in-poles $\alpha \beta w^{\pm1} q^{(\pm n_w-n_y)/2}$ and the out-poles $w^{\pm1} q^{(\pm n_w-n_y)/2}/\alpha \beta$ approach each other and pinch the contour $\mathbb{T}$ for $n_y=\pm n_w$. For these values of the fluxes, the residues $J(w^{\pm1})$ diverge as $((\alpha\beta)^2;q)_\infty^{-1}$, and the divergence compensates the zero of the prefactor. For all other values of the fluxes, the residues stay finite and their contributions vanish. As a result, we obtain\footnote{In order to cancel out many terms we used the equality \cite{Dimofte:2011py,Gahramanov:2015cva}
		\begin{equation}
			\frac{ (q^{1-m/2}z^{-1} ;q)_\infty} {( q^{-m/2} z;q)_\infty }
			=\frac{q^{m/2}}{(-z)^m} \frac{ (q^{1+m/2}z^{-1} ;q)_\infty} {( q^{m/2} z;q)_\infty }, \quad m\in {\mathbb Z}.
			\label{identity}\end{equation}}
	\begin{align} \nonumber
		\int_{\mathbb{T}}
		f(y)I_E \frac{dy}{2 \pi \textup{i} y}
		& = \left(\frac{a_5}{a_6}\right)^{n_5} \frac{(q^{1\pm n_\alpha} \alpha^{\mp 2};q)_\infty}{(q^{\pm n_\alpha} \alpha^{\pm 2};q)_\infty} \frac{(q^{1 \pm n_w} w^{ \mp 2};q)_\infty}{(q^{\pm n_w} w^{\pm 2};q)_\infty} \\ \label{prev}
		& \quad \times \left( f(w)\, \delta_{n_y,n_w}
		+ f(w^{-1})\, \delta_{n_y,-n_w} \right) .
	\end{align}
The factor $(a_5/a_6)^{n_5}$ has a simple origin. Indeed, for $a_5a_6=q$ and $n_6=-n_5$ the contribution of the fifth and sixth chiral doublets to the integrand of (\ref{electric}) reduces, by the identity (\ref{identity}), to the monomial $(a_5/a_6)^{n_5} z^{2m}$. Therefore, in this limit $I_E$ becomes $(a_5/a_6)^{n_5}$ times the index of the theory with $N_f=4$ chiral doublets, whose integrand contains the factor $z^{-4m}$ instead of $z^{-6m}$.

Now we denote $y=e^{2\pi\textup{i} \theta}$, $w=e^{2\pi\textup{i} \chi}$ and pass to the integration over the real variable $\theta$. Because of the arbitrariness of the function $f(y)$, we can give the index a distributional sense and rewrite the expression (\ref{prev}) in the following form
	\begin{align} \nonumber
		\sum_{m \in {\mathbb Z}} \oint \frac{(q^{1+\frac{n_\alpha\pm n_w\pm m}{2}} \alpha^{-1} w^{\mp} z^{\mp};q)_\infty}{(q^{\frac{n_\alpha \pm n_w \pm m}{2}} \alpha w^\pm z^\pm;q)_\infty}  \frac{(q^{1+\frac{-n_\alpha\pm n_y\pm m}{2}} \alpha y^{\mp} z^{\mp};q)_\infty}{(q^{\frac{-n_\alpha \pm n_y \pm m}{2}} \alpha^{-1} y^\pm z^\pm;q)_\infty} \frac{(1-q^m z^{\pm 2})}{q^m z^{4m}} \frac{dz}{4\pi i z} \\ \label{mainresult}
		= \frac{(q^{1\pm n_\alpha} \alpha^{\mp 2};q)_\infty}{(q^{\pm n_\alpha} \alpha^{\pm 2};q)_\infty} \frac{(q^{1 \pm n_w} w^{ \mp 2};q)_\infty}{(q^{\pm n_w} w^{\pm 2};q)_\infty} \left(\delta(\theta-\chi) \delta_{n_y,n_w}+\delta(\theta+\chi)\delta_{n_y,-n_w}\right) ,
	\end{align}
where $\delta(\theta)$ is the periodic Dirac delta function, $\delta(\theta+1)=\delta(\theta)$. The appearance of the Dirac delta functions describes the breaking of the flavor group $SU(4)$ to $Sp(4)$. Indeed, on the support of the delta functions the fugacities of the four doublets take the form $\{x_1^{\pm1}, x_2^{\pm1}\}$ with $x_1=\alpha w$ and $x_2=\alpha w^{-1}$, which parametrize the maximal torus of $Sp(4)$, and the dependence on the two independent fugacities $\alpha$ and $w$ reflects the rank of the unbroken group. In the parametrization (\ref{pars1}) only the subgroup $SU(2)_{diag}\times U(1)_\alpha \subset Sp(4)$ is manifest.

The coefficient in front of the delta functions also has a natural interpretation. The six mesons $M_{ij}=Q_iQ_j$ of the $N_f=4$ theory form the antisymmetric representation $\mathbf{6}$ of $SU(4)$, which decomposes as $\mathbf{5}\oplus\mathbf{1}$ under $Sp(4)$. The quantum constraint fixes the singlet, and the remaining five components are the Goldstone multiplets of the coset $SU(4)/Sp(4)$. On the support of the delta functions the charged components of the $\mathbf 5$ have the fugacities $x_1^{\pm1}x_2^{\pm1}=\alpha^{\pm2}$ and $x_1^{\pm1}x_2^{\mp1}=w^{\pm2}$, and their contributions are exactly the four factors on the right-hand side of (\ref{mainresult}).

The equality (\ref{mainresult}) has an interpretation as the inversion relation\footnote{The inversion relation is a simpler relation than the star-triangle relation within the framework of integrable lattice spin models of statistical mechanics, see, e.g. \cite{Stroganov:1979et,Baxter:1982xp,Jaekel:1982tp,Perk:1986nr}. In the context of the S-matrix for particle physics in $1+1$ dimensions it has the meaning of the unitarity relation.} for the exactly solvable lattice spin model introduced in \cite{Gahramanov:2015cva,Kels:2015bda}. In particular, this observation can be used \cite{Gahramanov:2022qge} for a better understanding of the correspondence \cite{Spiridonov:2010em,Gahramanov:2017ysd,Yamazaki:2018xbx,Yagi:2015lha} between supersymmetric gauge theories and integrable models of statistical mechanics.

\section{Concluding remarks}

We have shown that, starting from the three-dimensional $\mathcal N=2$ $s$-confining theory with six chiral doublets, one can obtain a confining theory with flavor symmetry breaking $SU(4)\to Sp(4)$ by analyzing the superconformal index. The index of the latter theory vanishes for generic fugacities and becomes a distribution supported on the maximal torus of the unbroken group. There are several potential directions to extend our work. It would be interesting to obtain the superconformal index with delta functions directly via the localization technique, if it is possible.

Although we studied only the $SU(2)$ supersymmetric gauge theory with $N_f=4$ chiral doublets, the approach discussed here also applies to gauge theories with other flavor groups. We would expect analogous behavior of the superconformal index for the general $SU(N)$ theory, however, it is computationally complicated to work with the multivariate analog of the basic hypergeometric integral identity (\ref{mainint}).

The four-dimensional $s$-confining theories are classified \cite{Csaki:1996sm,Csaki:1998th,Csaki:1998fm}. In the $SU(N_c)$ case, the chiral symmetry is always broken in such theories if one reduces the number of flavors to $N_c$. The corresponding three-dimensional analysis \cite{Csaki:2014cwa,Nii:2018wwj,Nii:2019ebv,Nii:2019dwi} for such theories needs to be performed.

Finally, there are many integrable lattice models \cite{Kels:2013ola,Gahramanov:2016ilb,Kels:2017vbc,Mullahasanoglu:2021xyf,de-la-Cruz-Moreno:2020xop,Dede:2022ofo} obtained recently via exact results from supersymmetric gauge theories. A similar approach can be used for obtaining the corresponding inversion relations. We hope to elaborate on this as well as on other topics elsewhere.

\section*{Acknowledgements}
The work originated from stimulating discussions with Vyacheslav Spiridonov, I thank him for his help, suggestions, and contributions to the ideas presented here. I would like to thank Keita Nii, Mustafa Mullahasanoglu, Erdal Catak, Yasin \c{S}ale, Merve Kaftan, and Ya\u{g}{\i}z Eren Kiremit for the discussions. I am grateful to the Institut des Hautes Etudes Scientifiques, IHES (Bures-sur-Yvette, France), for the warm hospitality, where part of this work was carried out. The visit to IHES was supported by the Nokia Fund in 2023.

\bibliographystyle{utphys}
\bibliography{references}

\end{document}